\documentclass[twocolumn,showpacs,preprintnumbers,amsmath,amssymb]{revtex4}
\usepackage{color}

\newcommand{\bea}{\begin{eqnarray}}
\newcommand{\eea}{\end{eqnarray}}

\begin{document}
\title{Axion electrodynamics and magnetohydrodynamics}
\author{Jai-chan Hwang${}^{1}$, Hyerim Noh${}^{2}$}
\address{${}^{1}$Center for Theoretical Physics of the Universe,
         Institute for Basic Science (IBS), Daejeon, 34051, Republic of Korea
         \\
         ${}^{2}$Theoretical Astrophysics Group, Korea Astronomy and Space Science Institute, Daejeon, 34055, Republic of Korea
         }


\begin{abstract}

We formulate axion-electrodynamics and magnetohydrodynamics (MHD) in the cosmological context assuming weak gravity. The two formulations are made for a general scalar field with general $f(\phi)$-coupling, and an axion as a massive scalar field with $\phi^2$-coupling, with the helical electromagnetic field. The $\alpha$-dynamo term appears naturally from the helical coupling in the MHD formulation. In the presence of the electromagnetic coupling, however, the Schr\"odinger and hydrodynamic formulations of the coherently oscillating axion are {\it not} available for the conventional $\phi$ coupling; instead, $\phi^2$ coupling allows successful formulations preserving the dark matter nature of the axion to nonlinear order. In the MHD formulation, direct couplings between the scalar and electromagnetic fields appear only for non-ideal MHD. We study gravitational and magnetic instabilities of the scalar field and axion MHDs.

\noindent

\end{abstract}
\maketitle


%
%
%
\section{Introduction}

The axion as a coherently oscillating massive scalar field is a cold or fuzzy dark matter (DM) candidate \cite{Hu-Barkana-Gruzinov-2000, Marsh-2016, Niemeyer-2020, Ferreira-2021, Hui-2021}. The axion electrodynamics (ED) is a central topic in experimental searches for the DM axion or axion-like particles in the laboratory \cite{Sikivie-1983, Irastorza-Redondo-2018, Sikivie-2021}. The pseudo-scalar nature of the axion allows a natural coupling with the helical electromagnetic (EM) field. The helical coupling can cause magnetic helicity generation, which has important implications in enhancing the large-scale magnetic field via dynamo action and inverse cascade \cite{Frisch-1975, Brandenburg-Subramanian-2005}. The origin and evolution of the magnetic field on the cosmic scale, unknown at the moment, are tied with the cosmological evolution. Thus, helically coupled axion is a subject of central importance in high-energy physics, astrophysics, and cosmology \cite{Grasso-Rubinstein-2001, Widrow-2002, Giovannini-2004, Durrer-Neronov-2013,  Subramanian-2016, Vachaspati-2021}.

Magnetohydrodynamics (MHD) is a convenient approximation for handling the EM field interacting with a conducting fluid. Here, we {\it aim} to provide the complete sets of equations for the ED and the MHD combined with a general scalar field and an axion as a massive scalar field, in the presence of additional coupling between the scalar field and the helical EM field. We consider the weak gravity limit in the cosmological context.

For a coherently oscillating axion, under the Klein transformation, we can derive the Schr\"odinger equation in the non-relativistic limit. Further applying the Madelung transformation, we have the quantum hydrodynamic equations revealing the nature of axion as the fuzzy (or cold) DM candidate \cite{Madelung-1927, Chavanis-2012, Hwang-Noh-2022a}. With the EM coupling, however, such transformations are {\it not} available for the conventional $\phi$-coupling commonly used in direct detection experiments of axion as the DM \cite{Sikivie-1983, Irastorza-Redondo-2018, Sikivie-2021}; for strong coupling, the DM nature of the axion is lost. Instead, a $\phi^2$-coupling allows successful transformations preserving the DM nature with coherent oscillation even to the nonlinear order.

We consider a scalar field generally coupled with the helical EM field. The Lagrangian density is
\bea
   & & L = {c^4 \over 16 \pi G} ( R - 2 \Lambda )
       - {1 \over 2} \phi^{;c} \phi_{,c} - V (\phi) + L_{\rm m}
   \nonumber \\
   & & \qquad
       - {1 \over 16 \pi} F_{ab} F^{ab}
       - {g_{\phi \gamma} \over 4} f (\phi)
       F_{ab} \widetilde F^{ab}
       + {1 \over c} J_{\rm e}^a A_a,
   \label{L}
\eea
where $L_{\rm m}$ is the fluid part, $R$ is the scalar curvature, $\Lambda$ is the cosmological constant, $\phi$ is the scalar field, $F_{ab}$ is the EM field strength tensor with $\widetilde F_{ab}$ its duel tensor, $J_{\rm e}^a$ is the electric four-current, and $A_a$ is the four-potential; $F_{ab} \widetilde F^{ab} = - 4 E^a B_a$ is parity-odd and leads to asymmetry between the two circular polarization states, thus helical.

Here, {\it assuming} weak gravity limit (see below) in cosmological context we will formulate the ED and MHD for a general scalar field with $V(\phi)$ and $f(\phi)$ {\it without} using the transformations, and for an axion with massive $V$ and $f = {1 \over 2} \phi^2$ {\it with} the transformations. These are the scalar field-ED and MHD in Sections \ref{SF-ED} and \ref{SF-MHD}, and axion-ED and MHD in Sections \ref{Axion-ED} and \ref{Axion-MHD}, respectively. In Section \ref{Instability} we investigate the gravitational and magnetic instabilities of the scalar field and axion MHDs. Section \ref{Discussion} is a discussion.

%
%
%
\section{Scalar field-ED}
                                           \label{SF-ED}

In this work, we consider a weak gravity limit of Einstein's gravity. The metric tensor convention is
\bea
   & & g_{00} = - \left( 1 + 2 {\Phi \over c^2} \right), \quad
       g_{0i} = - a {P_i \over c^3},
   \nonumber \\
   & & g_{ij} = a^2 \left( 1 - 2 {\Psi \over c^2} \right),
\eea
where $x^0 \equiv c t$ and $a(t)$ is the cosmic scale factor. As the weak gravity limit, we {\it assume}
\bea
   & & {\Phi \over c^2} \ll 1, \quad
       {\Psi \over c^2} \ll 1,
\eea
thus keep only to linear order in metric perturbation. In the current cosmological paradigm, we have $\Phi/c^2 \sim 10^{-5}$ or less in observed cosmological scales, thus indeed sufficiently small. However, we will keep the EM and the scalar fields fully relativistic and nonlinear, and this is why we keep two different potentials $\Phi$ and $\Psi$. Furthermore, in the weak gravity limit, the $g_{0i}$-component is non-vanishing \cite{Hwang-Noh-2016}. A consistent weak gravity limit combined with the relativistic matter parts is available in the uniform-expansion gauge, setting the expansion scalar of the normal frame vector $\theta \equiv n^a_{\;\; ;a}$ (which is minus of the trace of extrinsic curvature, $-K^i_i$) uniform in space; this differs from the zero-shear gauge setting the transverse part of $P_i$ equal to zero as the temporal gauge condition \cite{Hwang-Noh-2016, Noh-Hwang-Bucher-2019}. Later, for simplicity, we will {\it assume} slow-motion ($v^i v_i/c^2 \ll 1$) limit for the fluid part; for the fluid conservation equations, we will further {\it assume} a nonrelativistic limit. But, the scalar field and EM fields are kept relativistic, and the whole formulation is nonlinear.

Maxwell's equations, in the normal (laboratory) frame of reference, are modified by the axion-coupling as
\bea
   & & \nabla \cdot {\bf E}
       = 4 \pi a ( \varrho_{\rm e} + \varrho_{\rm e\phi} ),
   \label{Maxwell-1-WG} \\
   & & {1 \over c}
       \left( a^2 {\bf E} \right)^{\displaystyle{\cdot}}
       = a \nabla \times {\bf B}
       - {4 \pi a^2 \over c} ( {\bf j}_{\rm e}
       + {\bf j}_{\rm e\phi} ),
   \label{Maxwell-2-WG} \\
   & & \nabla \cdot {\bf B} = 0,
   \label{Maxwell-3-WG} \\
   & & {1 \over c}
       \left( a^2 {\bf B} \right)^{\displaystyle{\cdot}}
       = - a \nabla \times {\bf E},
   \label{Maxwell-4-WG}
\eea
with the axion-induced electric charge and current densities, respectively \cite{Sikivie-1983, Wilczek-1987}
\bea
   \varrho_{\rm e\phi}
       = - g_{\phi \gamma} {1 \over a}
       {\bf B} \cdot \nabla f, \quad
       {\bf j}_{\rm e\phi}
       = g_{\phi \gamma} \left( {\bf B} \dot f
       - {c \over a} {\bf E} \times \nabla f \right).
   \label{rho-j-phi}
\eea
These are Gauss's law, Amp$\grave{\rm e}$re's law, no monopole condition, and Faraday's law, respectively, and are valid in the weak gravity limit. We use the Gaussian unit \cite{Jackson-1975}.

The Klein-Gordon equation gives \cite{Sikivie-1983}
\bea
   \ddot \phi + 3 {\dot a \over a} \dot \phi
       - c^2 {\Delta \over a^2} \phi
       + ( c^2 + 2 \Phi) V_{,\phi}
       = c^2
       g_{\phi \gamma} f_{,\phi} {\bf E} \cdot {\bf B}.
   \label{EOM-WG}
\eea
We kept the gravitational potential $\Phi$ in the weak gravity limit, as the mass term in the scalar field potential $V$ is already $c^2$ order; for a massive field, $V = {1 \over 2} {m^2 c^2 \over \hbar^2} \phi^2$. We have $F_{ab} \widetilde F^{ab} = - 4 {\bf E} \cdot {\bf B}$ which is related to the time derivative of the magnetic helicity, $\int_V {\bf A} \cdot {\bf B} d^3 x$ with ${\bf B} \equiv \nabla \times {\bf A}$ \cite{Blackman-2015}.

For the fluid, for simplicity, we consider only the continuity and Euler equations in the non-relativistic limit
\bea
   & & \dot \varrho + 3 {\dot a \over a} \varrho
       + {1 \over a} \nabla \cdot
       \left( \varrho {\bf v} \right)
       = 0,
   \label{Continuity-fluid} \\
   & & \dot {\bf v} + {\dot a \over a} {\bf v}
       + {1 \over a} {\bf v} \cdot \nabla {\bf v}
       + {1 \over a} \nabla \Phi
       + {1 \over \varrho a} \left( \nabla p
       + \nabla_j \Pi^j_i \right)
   \nonumber \\
   & & \qquad
       = {1 \over \varrho} \left(
       \varrho_{\rm e} {\bf E}
       + {1 \over c} {\bf j}_{\rm e} \times {\bf B} \right),
   \label{Euler-fluid}
\eea
where $\varrho$, ${\bf v}$, $p$ and $\Pi_{ij}$ are the density, velocity, pressure and anisotropic stress, respectively. The right-hand side of Eq.\ (\ref{Euler-fluid}) is the Lorentz force. The $g_{\phi \gamma}$-coupling couples the EM field with the scalar field only, and does not directly affect the fluid conservation equations.

For the gravity, we have \cite{Noh-Hwang-Bucher-2019}
\bea
   & & {\Delta \over a^2} \Phi
       = {4 \pi G \over c^2} \bigg( \mu + 3 p
       + {2 \over c^2} \dot \phi^2 - 2 V
       + {E^2 + B^2 \over 4 \pi} \bigg)
   \nonumber \\
   & & \qquad
       + 3 {\ddot a \over a} - \Lambda c^2,
   \label{Poisson-eq} \\
   & & {\Delta \over a^2} \Psi
       = {4 \pi G \over c^2} \bigg( \mu
       + {1 \over 2 c^2} \dot \phi^2 + V
       + {1 \over 2 a^2} \phi^{,i} \phi_{,i}
       + {E^2 + B^2 \over 8 \pi} \bigg)
   \nonumber \\
   & & \qquad
       - {3 \over 2} {\dot a^2 \over a^2}
       + {\Lambda c^2 \over 2},
   \label{Poisson-eq-Psi}
\eea
thus $\Psi \neq \Phi$ in our weak gravity approximation. These were derived in the uniform-expansion gauge; in the zero-shear gauge, $3 p$-term in Eq.\ (\ref{Poisson-eq}) is missing, which contradicts the exact result known in the spherically symmetric system \cite{Hwang-Noh-2016}. We kept fluid variables (energy density $\mu \equiv \varrho c^2$ and pressure $p$) to the weak gravity and slow-motion limits; in the non-relativistic limit we can further ignore the pressure term. We note that the EM and the scalar fields are still fully relativistic. The $g_{\phi \gamma}$-coupling term does not directly appear in gravity because it does not contribute to the energy-momentum tensor. Notice that only $\Phi$ (not $\Psi$) couples with the fields and the fluid; this is because we consider the non-relativistic order in the fluid. For weak gravity limit combined with fully relativistic matter and conventional EM field, see \cite{Hwang-Noh-2016, Noh-Hwang-Bucher-2019}.

These complete the scalar field-ED in the weak gravity combined with the non-relativistic fluid: the complete set is Eqs.\ (\ref{Maxwell-1-WG})-(\ref{Poisson-eq}). We considered general $V(\phi)$ and $f(\phi)$ in the cosmological context. In the Friedmann background, Eqs.\ (\ref{EOM-WG}), (\ref{Continuity-fluid}), (\ref{Poisson-eq}) and (\ref{Poisson-eq-Psi}) include the background order equations, and in perturbation analysis we may subtract the background order, see Section \ref{Instability}. The Friedmann background cannot accommodate the EM field. By setting $a \equiv 1$ and $\Lambda = 0$, we recover equations in the Minkowski background.

Our basic equations in (\ref{Maxwell-1-WG})-(\ref{Poisson-eq-Psi}) are derived from the fully relativistic formulation by sequentially taking the weak gravity, slow-motion and non-relativistic limits; the latter two limits applied only to the fluid component. A fully relativistic extension of the present work is currently in progress \cite{Hwang-Noh-2022b}.

%
%
%
\section{Scalar field-MHD}
                                           \label{SF-MHD}

The {\it MHD approximation} considers $4 \pi \sigma T \gg 1$ where $\sigma$ is the electrical conductivity and $T$ is the characteristic time scale of variation of the EM fields \cite{Somov-1994}. In the slow-motion limit, MHD (i) {\it adopts} a simple form of Ohm's law as
\bea
   {\bf j}_e = \sigma \left( {\bf E}
       + {1 \over c} {\bf v} \times {\bf B} \right),
   \label{Ohms-law}
\eea
(ii) the displacement current term in Amp$\grave{\rm e}$re's law is negligible, (iii) $E^2$ term in Eq.\ (\ref{Poisson-eq}) is negligible, and (iv) in ordinary MHD, $\varrho_{\rm e} {\bf E}$ is negligible compared with the second Lorentz force term, but not in the presence of the $g_{\phi \gamma}$-coupling; using the Ohm's law, we can show that only $\nabla \cdot {\bf E}$ term coming from $\varrho_{\rm e} {\bf E}$ using Gauss' law is negligible compared with the second Lorentz force term [see Eq.\ (\ref{Euler-fluid-MHD})].

In the MHD, the fundamental dynamic variables are hydrodynamic ones like $\varrho$, ${\bf v}$, and the magnetic field ${\bf B}$; in our case, we have additional scalar field variable $\phi$ and gravity $\Phi$. The Ohm's law determines ${\bf E}$, Gauss' law determines $\varrho_{\rm e}$, and Amp$\grave{\rm e}$re's law determines ${\bf j}_{\rm e}$; these are respectively,
\bea
   & & {\bf E} = - {1 \over c} {\bf v} \times {\bf B}
       + {{\bf j}_{\rm e} \over \sigma}, \quad
       \varrho_{\rm e} = {\nabla \cdot {\bf E} \over 4 \pi a}
       - \varrho_{\rm e\phi},
   \nonumber \\
   & &
       {\bf j}_{\rm e} = {c \nabla \times {\bf B} \over 4 \pi a}
       - {\bf j}_{\rm e\phi}.
   \label{Ohm-Gauss-Ampere-MHD}
\eea
Notice that in our case with $g_{\phi \gamma}$-coupling, ${\bf E}$ and ${\bf j}_{\rm e}$ are coupled; we may truncate it (using smallness of either $1/\sigma$ or $g_{\phi \gamma}$) at some point depending on the situation.

Using the Ohm's law and Amp$\grave{\rm e}$re's law in Eq.\ (\ref{Ohm-Gauss-Ampere-MHD}), and assuming constant $\sigma$, the Faraday equation gives
\bea
   & & {1 \over a^2}
       \left( a^2 {\bf B} \right)^{\displaystyle{\cdot}}
       - {1 \over a} \nabla \times ( {\bf v} \times {\bf B} )
       - {c^2 \Delta {\bf B} \over 4 \pi \sigma a^2}
   \nonumber \\
   & & \qquad
       = {c g_{\phi \gamma} \over \sigma a}
       \nabla \times \left( \dot f {\bf B}
       - {c \over a} {\bf E} \times \nabla f \right).
   \label{Faraday-MHD}
\eea
The second term in the left-hand side is the flux conserving induction term (in the absence of the other terms the magnetic field is frozen-in with the fluid), and the third term is the diffusion. The right-hand side can work as the scalar field source for the magnetic field; especially, the first term can work as the $\alpha$-dynamo \cite{Krause-Radler-1980} with $\alpha = c g_{\phi \gamma} \dot f / \sigma$ \cite{Field-Carroll-2000, Campanelli-Giannotti-2005}. For an ideal MHD ($\sigma \rightarrow \infty$), the $g_{\phi \gamma}$-coupling disappears. One remaining equation is $\nabla \cdot {\bf B} = 0$. Linear solutions with exponential growth due to the $\alpha$-dynamo term are given in Eq.\ (\ref{B-MHD-solutions}). 

For the fluid, the continuity equation in (\ref{Continuity-fluid}) remains the same. The Euler equation in (\ref{Euler-fluid}), using the Gauss' and Amp$\grave{\rm e}$re's laws in Eq.\ (\ref{Ohm-Gauss-Ampere-MHD}), gives
\bea
   & & \dot {\bf v} + {\dot a \over a} {\bf v}
       + {1 \over a} {\bf v} \cdot \nabla {\bf v}
       + {1 \over a} \nabla \Phi
       + {1 \over \varrho a} \left( \nabla p
       + \nabla_j \Pi^j_i \right)
   \nonumber \\
   & & \qquad
       = {(\nabla \times {\bf B}) \times
       {\bf B} \over 4 \pi \varrho a}
       + g_{\phi \gamma} {1 \over \varrho a}
       {\bf E} \cdot {\bf B} \nabla f.
   \label{Euler-fluid-MHD}
\eea
Notice the presence of $g_{\phi \gamma}$-coupling contribution in the scalar field-MHD while such a term is absent in the original scalar field-ED in Eq.\ (\ref{Euler-fluid}). Using the Amp$\grave{\rm e}$re's and Ohm's laws in Eq.\ (\ref{Ohm-Gauss-Ampere-MHD}), we have
\bea
   {\bf E} \cdot {\bf B}
       = {c {\bf B} \cdot ( \nabla \times {\bf B} )
       \over 4 \pi \sigma a}
       - {g_{\phi \gamma} \over \sigma}
       \left[ \dot f B^2
       - {c \over a} {\bf B} \cdot
       ( {\bf E} \times \nabla f ) \right].
   \label{E-dot-B-MHD}
\eea
For ${\bf E}$-term in the right-hand side, we may again use the Ohm's law with a truncation, see below Eq.\ (\ref{Ohm-Gauss-Ampere-MHD}). For the scalar field, Eq.\ (\ref{EOM-WG}) remains the same and we only need ${\bf E} \cdot {\bf B}$ expressed in the MHD approximation as in Eq.\ (\ref{E-dot-B-MHD}). In the ideal MHD limit, we have ${\bf E} \cdot {\bf B} = 0$, and the $g_{\phi \gamma}$-coupling {\it entirely} disappears. Thus, for ideal MHD with $\sigma \rightarrow \infty$, the scalar field and magnetic field are coupled only through gravity. For the gravity, Eq.\ (\ref{Poisson-eq}) remains the same except for the absence of $E^2$-term.

These complete the MHD approximation coupled with a scalar field with general $V(\phi)$ and $f(\phi)$: the complete set is Eqs.\ (\ref{Ohm-Gauss-Ampere-MHD})-(\ref{E-dot-B-MHD}) together with Eqs.\ (\ref{EOM-WG}), (\ref{Continuity-fluid}) and (\ref{Poisson-eq}) for the scalar field, fluid and gravity, respectively.

%
%
%
\section{Axion-ED}
                                           \label{Axion-ED}

%
%
%
\subsection{Klein transformation}

From now on, we consider a massive field {\it with} $f = {1 \over 2} \phi^2$ and call it axion. We will use two transformations that lead to the Schr\"odinger formulation and Madelung's hydrodynamic formulation. In other forms of $f$-coupling (including the conventional $f = \phi$ coupling) with sufficiently large coupling strength $g_{\phi\gamma}$, it is {\it difficult} to apply the two transformations with time-average. In such cases, we can use the ED and MHD formulations made for general $V (\phi)$ and $f (\phi)$ in the previous two sections.

The {\it Klein transformation} is \cite{Klein-1926, Chavanis-Matos-2017}
\bea
   \phi ({\bf x}, t)
       \equiv {\hbar \over \sqrt{2m}}
       \left[ \psi ({\bf x}, t) e^{- i \omega_c t}
       + \psi^* ({\bf x}, t) e^{+ i \omega_c t} \right],
   \label{Klein-tr}
\eea
where $\phi$ is a real scalar field, and $\psi$ is a complex wave function; $\omega_c \equiv m c^2/\hbar$ is the Compton frequency. This {\it ansatz} is valid if the scalar field oscillates with Compton frequency. On sub-Compton scale, the Laplacian term in Eq.\ (\ref{EOM-WG}) dominates and the scalar field does not oscillate. Thus, the Klein transformation works only on super-Compton scale \cite{Hwang-Noh-2022a}.

{\it Ignoring} the rapidly oscillating parts (by taking time average), we have $f = \hbar^2 |\psi|^2/(2 m)$, and Eq.\ (\ref{rho-j-phi}) gives
\bea
   & & \varrho_{\rm e\phi}
       = - {\hbar^2 g_{\phi \gamma} \over 2 m} {1 \over a}
       {\bf B} \cdot \nabla |\psi|^2,
   \nonumber \\
   & & {\bf j}_{\rm e\phi}
       = {\hbar^2 g_{\phi \gamma} \over 2 m}
       \left[ {\bf B} ( |\psi|^2 )^{\displaystyle{\cdot}}
       - {c \over a} {\bf E} \times \nabla |\psi|^2 \right].
   \label{rho-j-phi-KT}
\eea
Equation (\ref{EOM-WG}), in the non-relativistic ($c \rightarrow \infty$) limit \cite{Hwang-Noh-2022a}, gives
\bea
   i \hbar \left( \dot \psi
       + {3 \over 2} {\dot a \over a} \psi \right)
       = - {\hbar^2 \over 2 m} {\Delta \over a^2} \psi
       + m \Phi \psi
       - {\hbar^2 g_{\phi \gamma} \over 2 m}
       {\bf E} \cdot {\bf B} \psi,
   \label{Schrodinger-eq}
\eea
which is the Schr\"odinger equation in expanding background, including the gravity and the EM coupling. For the gravity, from Eq.\ (\ref{Poisson-eq}), we have
\bea
   {\Delta \over a^2} \Phi
       = 4 \pi G \left( \varrho + m |\psi|^2
       + {E^2 + B^2 \over 4 \pi c^2} \right)
       + 3 {\ddot a \over a} - \Lambda c^2,
   \label{Poisson-eq-Klein}
\eea
where we ignored (by time-average) oscillating parts and took the non-relativistic limit for the axion; we have {\it not} imposed the non-relativistic condition in the EM part.

%
%
%
\subsection{Madelung transformation}

Assuming the first and second derivatives of $u$ are well defined, under the {\it Madelung transformation} \cite{Madelung-1927}
\bea
   \psi \equiv \sqrt{\varrho_\phi \over m}
       e^{i m u/\hbar},
   \label{Madelung-tr}
\eea
Eq.\ (\ref{rho-j-phi-KT}) gives
\bea
   & & \varrho_{\rm e\phi}
       = - {\hbar^2 g_{\phi \gamma} \over 2 m^2} {1 \over a}
       {\bf B} \cdot \nabla \varrho_\phi,
   \nonumber \\
   & & {\bf j}_{\rm e\phi}
       = {\hbar^2 g_{\phi \gamma} \over 2 m^2}
       \left( {\bf B} \dot \varrho_\phi
       - {c \over a} {\bf E} \times \nabla \varrho_\phi \right).
   \label{rho-j-phi-MT}
\eea
Imaginary and real parts, respectively, of Eq.\ (\ref{Schrodinger-eq}) give \cite{Madelung-1927, Chavanis-2012, Hwang-Noh-2022a}
\bea
   & & \dot \varrho_\phi + 3 {\dot a \over a} \varrho_\phi
       + {1 \over a} \nabla \cdot
       \left( \varrho_\phi {\bf v}_\phi \right)
       = 0,
   \label{Continuity-axion} \\
   & & \dot {\bf v}_\phi + {\dot a \over a} {\bf v}_\phi
       + {1 \over a} {\bf v}_\phi \cdot \nabla {\bf v}_\phi
       + {1 \over a} \nabla \Phi
   \nonumber \\
   & & \qquad
       = {\hbar^2 \over 2 m^2} {1 \over a^3} \nabla \left(
       {\Delta \sqrt{\varrho_\phi}
       \over \sqrt{\varrho_\phi}} \right)
       + {\hbar^2 g_{\phi \gamma} \over 2 m^2 a}
       \nabla ( {\bf E} \cdot {\bf B} ),
   \label{Euler-axion}
\eea
where we identified ${\bf v}_\phi \equiv {1 \over a} \nabla u$, thus $\nabla \times {\bf v}_\phi = 0$. The potential-flow nature of the axion velocity is an important characteristic of the axion fluid; the quantized vortices (see below) appear in the Schr\"odinger formulation in Eq.\ (\ref{Schrodinger-eq}), and their cosmological roles are studied in \cite{Rindler-Dallas-Shapiro-2012, Alexander-2021, Hui-Joyce-2021}. The first term in the right-hand side of Eq.\ (\ref{Euler-axion}) is the quantum stress \cite{Takabayasi-1952}. Notice the difference in the EM parts between Eqs.\ (\ref{Euler-fluid}) and (\ref{Euler-axion}).

The non-equivalence between the Schr\"odinger formulation and the hydrodynamic formulation by Madelung is recognized in the literature: while the hydrodynamic formulation has potential flow without vortex, the Schr\"odinger formulation has quantized vortices \cite{Takabayasi-1952, Wallstrom-1994}. The single valuedness of the wave function demands the circulation around any closed path to be quantized
\bea
   \Gamma = \oint_C a {\bf v} \cdot d \ell
       = \oint_C (\nabla u) \cdot d \ell
       = \oint_C du = n {h \over m},
\eea
where $n$ is an integer and $n \neq 0$ for a path encircling vanishing wavefunction \cite{Hirschfelder-1974}. Using Stokes' theorem, the circulation is related to the vorticity $\vec{\omega} \equiv {1 \over a} \nabla \times {\bf v}$ as
\bea
   \Gamma
       = \int \hskip -.3cm \int_S a ( \nabla \times {\bf v} )
       \cdot d \vec{S}
       = \int \hskip -.3cm \int_S a^2 \vec{\omega}
       \cdot d \vec{S}.
\eea

The hydrodynamic formulation reveals the fuzzy DM nature of axion preserved for $\phi^2$-coupling; for $\phi$-coupling, however, these two transformations are {\it not} possible, and for a sufficiently large coupling strength $g_{\phi\gamma}$ the DM nature is {\it lost} to nonlinear order, see later. For the gravity, Eq.\ (\ref{Poisson-eq-Klein}) remains the same with $\varrho_\phi = m |\psi|^2$.

Combining with the fluid equations in (\ref{Continuity-fluid}) and (\ref{Euler-fluid}) and Maxwell's equations in (\ref{Maxwell-1-WG})-(\ref{rho-j-phi}) we have the complete sets of axion-ED in either the Schr\"odinger formulation or the fluid formulation for the axion field.

%
%
%
\section{Axion-MHD}
                                           \label{Axion-MHD}

Now, we present the {\it axion-MHD} approximation for $\phi^2$-coupling. The MHD conditions in Eqs.\ (\ref{Ohms-law}) and (\ref{Ohm-Gauss-Ampere-MHD}) remain the same. Using the Ohm's law and Amp$\grave{\rm e}$re's law in Eq.\ (\ref{Ohm-Gauss-Ampere-MHD}), and assuming constant $\sigma$, the Faraday equation gives
\bea
   & & {1 \over a^2}
       \left( a^2 {\bf B} \right)^{\displaystyle{\cdot}}
       - {1 \over a} \nabla \times ( {\bf v} \times {\bf B} )
       - {c^2 \Delta {\bf B} \over 4 \pi a^2 \sigma}
   \nonumber \\
   & & \qquad
       = {c \hbar^2 g_{\phi \gamma} \over 2 m^2 \sigma a}
       \nabla \times \left( \dot \varrho_\phi {\bf B}
       - {c \over a} {\bf E} \times \nabla \varrho_\phi \right).
   \label{Faraday-MHD-axion}
\eea
The first term in the right-hand side works as the $\alpha$-effect of mean field dynamo with $\alpha = {c \hbar^2 g_{\phi \gamma} \over 2 m^2 \sigma} \dot \varrho_\phi$. In dynamo theory, the $\alpha$-term arises from the induction term using the mean field MHD \cite{Krause-Radler-1980, Moffatt-1983, Choudhuri-1998}; kinetic energy is converted to the magnetic one by turbulent motion. Here the $g_{\phi \gamma}$-coupling directly causes the $\alpha$-term for a finite $\sigma$. Linear solutions with exponential growth are given in Eq.\ (\ref{B-MHD-solutions}).

For the fluid, the continuity equation in (\ref{Continuity-fluid}) remains the same. The Euler equation in (\ref{Euler-fluid}), using the Gauss' and Amp$\grave{\rm e}$re's laws in Eq.\ (\ref{Ohm-Gauss-Ampere-MHD}), gives
\bea
   & & \dot {\bf v} + {\dot a \over a} {\bf v}
       + {1 \over a} {\bf v} \cdot \nabla {\bf v}
       + {1 \over a} \nabla \Phi
       + {1 \over \varrho a} \left( \nabla p
       + \nabla_j \Pi^j_i \right)
   \nonumber \\
   & & \qquad
       = {(\nabla \times {\bf B}) \times
       {\bf B} \over 4 \pi \varrho a}
       + {\hbar^2 g_{\phi \gamma} \over 2 m^2 \varrho a}
       {\bf E} \cdot {\bf B} \nabla \varrho_\phi.
   \label{Euler-fluid-MHD-axion}
\eea
Using the Amp$\grave{\rm e}$re's and Ohm's laws in Eq.\ (\ref{Ohm-Gauss-Ampere-MHD}), we have
\bea
   & & {\bf E} \cdot {\bf B}
       = {c {\bf B} \cdot ( \nabla \times {\bf B} )
       \over 4 \pi \sigma a}
   \nonumber \\
   & & \qquad
       - {\hbar^2 g_{\phi \gamma} \over 2 m^2 \sigma}
       \left[ B^2 \dot \varrho_\phi
       - {c \over a} {\bf B} \cdot
       ( {\bf E} \times \nabla \varrho_\phi ) \right].
   \label{E-dot-B-MHD-axion}
\eea

These complete the axion-MHD with $\phi^2$-coupling: the complete set is Eqs.\ (\ref{Faraday-MHD-axion})-(\ref{E-dot-B-MHD-axion}) together with Eqs.\ (\ref{Ohm-Gauss-Ampere-MHD}) and  (\ref{Continuity-fluid}) for the EM field and fluid. For the gravity, Eq.\ (\ref{Poisson-eq-Klein}) is valid without the $E^2$ term. For the axion, we have either the Schr\"odinger formulation in Eq.\ (\ref{Schrodinger-eq}) or the Madelung's hydrodynamic formulation in Eqs.\ (\ref{Continuity-axion}) and (\ref{Euler-axion}), with ${\bf E} \cdot {\bf B}$ in Eq.\ (\ref{E-dot-B-MHD-axion}). In the ideal MHD, the $g_{\phi \gamma}$-coupling effect {\it entirely} disappears.

%
%
%
\section{Instabilities of axion-MHD}
                                           \label{Instability}

%
%
%
\subsection{Gravitational instability}

As an application, we consider gravitational instability of the fluid and axion system caused by the MHD with helical $\phi^2$-coupling. We set $\varrho \rightarrow \varrho + \delta \varrho \equiv \varrho (1 + \delta)$, and similarly for $p$ and $\varrho_\phi$ . We keep to the linear perturbation orders in the fluid and the axion but keep nonlinear order in the magnetic field; this is because EM fields always appear in quadratic (thus nonlinear) combinations. To be consistent, we have to expand the fluid and axion field at least to the second-order as well, but here for simplicity, we ignore writing these nonlinear terms. To the background order, Eqs.\ (\ref{Continuity-fluid}), (\ref{Continuity-axion}) and (\ref{Poisson-eq-Klein}) give
\bea
   (a^3 \varrho)^{\displaystyle{\cdot}} = 0 =
       (a^3 \varrho_\phi)^{\displaystyle{\cdot}}, \quad
       {\ddot a \over a}
       = - {4 \pi G \over 3} ( \varrho + \varrho_\phi )
       + {\Lambda c^2 \over 3}.
\eea
For perturbed parts, we subtract the background equations.

For the fluid perturbation, Eqs.\ (\ref{Continuity-fluid}) and (\ref{Euler-fluid}) give
\bea
   & & \dot \delta + {1 \over a} \nabla \cdot {\bf v} = 0,
   \label{dot-delta-eq-fluid} \\
   & & \dot {\bf v} + {\dot a \over a} {\bf v}
       + {1 \over a} \nabla \Phi
       + {1 \over \varrho a} \left( \nabla \delta p
       + \nabla_j \Pi^j_i \right)
   \nonumber \\
   & & \qquad
       = {(\nabla \times {\bf B}) \times
       {\bf B} \over 4 \pi \varrho a}
       + {\hbar^2 g_{\phi \gamma} \over 2 m^2 a}
       {\varrho_\phi \over \varrho}
       {\bf E} \cdot {\bf B} \nabla \delta_\phi.
   \label{v-eq-fluid-linear}
\eea
Keeping nonlinear order only in EM field, Eq.\ (\ref{E-dot-B-MHD}) gives
\bea
   {\bf E} \cdot {\bf B}
       = {c {\bf B} \cdot ( \nabla \times {\bf B} )
       \over 4 \pi \sigma a}
       - {\hbar^2 g_{\phi \gamma} \over 2 m^2 \sigma}
       B^2 \dot \varrho_\phi.
   \label{E-dot-B-MHD-linear}
\eea
By taking divergence and curl operations, we have
\bea
   & & \ddot \delta + 2 {\dot a \over a} \dot \delta
       - {\Delta \over a^2} \Phi
       - {1 \over \varrho a^2} \left( \Delta \delta p
       + \nabla_i \nabla_j \Pi^{ij} \right)
   \nonumber \\
   & & \qquad
       = - {\nabla \cdot [ (\nabla \times {\bf B}) \times
       {\bf B} ] \over 4 \pi \varrho a^2},
   \label{delta-eq-MHD} \\
   & & {1 \over a^2} ( a^2 \vec{\omega}
       )^{\displaystyle{\cdot}}
       + {1 \over \varrho a^2} \eta_{ijk}
       \nabla^j \nabla_\ell \Pi^{k \ell}
       = {\nabla \times [ (\nabla \times {\bf B}) \times
       {\bf B} ] \over 4 \pi \varrho a^2},
   \label{omega-eq-MHD}
\eea
where $\vec{\omega} \equiv {1 \over a} \nabla \times {\bf v}$; we ignore $g_{\phi \gamma}$ contribution, the last term in Eq.\ (\ref{v-eq-fluid-linear}), which already involves perturbed axion density in a nonlinear context. Thus, the ideal MHD can source the density and the angular momentum \cite{Wasserman-1978, Kim-Olinto-Rosner-1996}. We note that, although we kept only to the linear order in perturbed fluid variables (by {\it ignoring} writing the nonlinear terms), what is generated by the magnetic field is {\it nonlinear} order fluid perturbations. The quadratic combinations of magnetic field source the density and rotational perturbations, and as the magnetic field is already a perturbed order, the quadratic combinations work as nonlinear source terms.

For the axion perturbation, Eqs.\ (\ref{Continuity-axion}) and (\ref{Euler-axion}) give
\bea
   & & \dot \delta_\phi
       + {1 \over a} \nabla \cdot {\bf v}_\phi = 0,
   \\
   & & \dot {\bf v}_\phi + {\dot a \over a} {\bf v}_\phi
       + {1 \over a} \nabla \Phi
       = {\hbar^2 \nabla \Delta \delta_\phi \over 4 m^2 a^3}
       + {\hbar^2 g_{\phi \gamma} \over 2 m^2 a}
       \nabla ({\bf E} \cdot {\bf B}),
\eea
By taking divergence and curl operations, we have
\bea
   & & \ddot \delta_\phi
       + 2 {\dot a \over a} \dot \delta_\phi
       - {\Delta \over a^2} \Phi
       + {\hbar^2 \Delta^2 \over 4 m^2 a^4} \delta_\phi
       = - {\hbar^2 g_{\phi \gamma} \Delta \over 2 m^2 a^2}
       {\bf E} \cdot {\bf B},
   \label{delta-eq-MHD-axion} \\
   & & {1 \over a^2} ( a^2 \vec{\omega}_\phi
       )^{\displaystyle{\cdot}} = 0,
   \label{omega-eq-MHD-axion}
\eea
where $\vec{\omega}_\phi \equiv {1 \over a} \nabla \times {\bf v}_\phi$. The $g_{\phi \gamma}$-coupling sources axion density perturbation for a finite $\sigma$, whereas the vorticity of the axion is free from the coupling due to the potential nature of ${\bf v}_\phi$. In hydrodynamic formulation, we have $\vec{\omega}_\phi = 0$ exactly to nonlinear order, see Eq.\ (\ref{Euler-axion}); as mentioned, to make the hydrodynamic formulation equivalent to the Schr\"odinger formulation we need additional quantized vortices added by hand \cite{Wallstrom-1994}. For gravity, Eq.\ (\ref{Poisson-eq-Klein}) gives
\bea
   {\Delta \over a^2} \Phi
       = 4 \pi G \left( \varrho \delta
       + \varrho_\phi \delta_\phi
       + {B^2 \over 4 \pi c^2} \right).
   \label{Poisson-eq-MHD-linear}
\eea

Combining Eqs.\ (\ref{delta-eq-MHD}), (\ref{delta-eq-MHD-axion}) and (\ref{Poisson-eq-MHD-linear}), ignoring the anisotropic stress, we have
\bea
   & & \ddot \delta + 2 {\dot a \over a} \dot \delta
       - 4 \pi G \left( \varrho \delta
       + \varrho_\phi \delta_\phi
       + {B^2 \over 4 \pi c^2} \right)
       - {\Delta \over a^2} {\delta p \over \varrho}
   \nonumber \\
   & & \qquad
       = - {\nabla \cdot [ (\nabla \times {\bf B}) \times
       {\bf B} ] \over 4 \pi \varrho a^2},
   \label{ddot-delta-MHD} \\
   & & \ddot \delta_\phi
       + 2 {\dot a \over a} \dot \delta_\phi
       - 4 \pi G \left( \varrho \delta
       + \varrho_\phi \delta_\phi
       + {B^2 \over 4 \pi c^2} \right)
       + {\hbar^2 \Delta^2 \over 4 m^2 a^4} \delta_\phi
   \nonumber \\
   & & \qquad
       = - {\hbar^2 g_{\phi \gamma} \Delta \over 2 m^2 a^2}
       {\bf E} \cdot {\bf B}.
   \label{ddot-delta-axion-MHD}
\eea
Considering pure fluid and pure axion in Eqs.\ (\ref{ddot-delta-MHD}) and (\ref{ddot-delta-axion-MHD}), respectively, and by comparing the gravity term with the pressure/stress term, we have the Jeans criterion dividing the gravity and pressure dominating scales. For fluid and axion, respectively, we have
\bea
   & & {k_{\rm J} \over a} = {\sqrt{4 \pi G \varrho} \over v_s},
       \quad
       {k_{\rm J \phi} \over a} = ( 6 \Omega_\phi )^{1/4}
       \sqrt{{m H \over \hbar}},
   \label{Jeans-scale}
\eea
where $\Delta = - k^2$, $v_s \equiv \sqrt{\delta p/\delta \varrho}$, $\Omega_\phi \equiv \varrho_\phi/\varrho_{\rm cr}$, $\varrho_{\rm cr} \equiv 3 H^2/(8 \pi G)$, and $H \equiv {\dot a \over a}$.

%
%
%
\subsection{Magnetic instability}

For magnetic field, to the linear order, Eq.\ (\ref{Faraday-MHD-axion}) gives
\bea
   {1 \over a^2}
       \left( a^2 {\bf B} \right)^{\displaystyle{\cdot}}
       - {c^2 \Delta {\bf B} \over 4 \pi a^2 \sigma}
       = {c \hbar^2 g_{\phi \gamma} \over 2 m^2 \sigma a}
       \dot \varrho_\phi
       \nabla \times {\bf B}.
   \label{Faraday-MHD-linear}
\eea
In the case of general scalar field, from Eq.\ (\ref{Faraday-MHD}) we have $\hbar^2 \varrho_\phi / (2 m^2) \rightarrow f$. In Fourier space with ${\bf B} ({\bf k}, t) = \int d^3 x e^{-i{\bf k} \cdot {\bf x}} {\bf B} ({\bf x}, t)$, and using the orthonormal helicity (circular polarization) basis \cite{Jackson-1975} $( \widehat {\bf e}_+, \widehat {\bf e}_-, \widehat {\bf e}_3 )$ with $\widehat {\bf e}_\pm \equiv ( \widehat {\bf e}_1 \pm i \widehat {\bf e}_2 ) / \sqrt{2}$, $\widehat {\bf e}_3 \equiv {{\bf k} / k}$, and ${\bf B} \equiv B_+ \widehat {\bf e}_+ + B_- \widehat {\bf e}_- + B_3 \widehat {\bf e}_3$, we have
\bea
   {1 \over a^2} ( a^2 B_\pm )^{\displaystyle{\cdot}}
       = {c^2 \over 4 \pi \sigma}
       \left( - {k^2 \over a^2}
       \pm {2 \pi \hbar^2 g_{\phi \gamma} \over m^2 c}
       {k \over a} \dot \varrho_\phi \right) B_\pm,
\eea
with solutions \cite{Field-Carroll-2000, Campanelli-Giannotti-2005, Long-Vachaspati-2015}
\bea
   B_\pm
       = B_{\pm i} {a_i^2 \over a^2} {\rm exp}
       \left[  \int^t_{t_i}
       {c^2 \over 4 \pi \sigma}
       \left( - {k^2 \over a^2}
       \pm {2 \pi \hbar^2 g_{\phi \gamma} \over m^2 c}
       {k \over a} \dot \varrho_\phi \right) dt \right],
   \label{B-MHD-solutions}
\eea
and $B_3$ pure decaying. The first term is diffusion damping. The second term causes exponential growth of the magnetic field for small enough $k$ and steady $\dot \varrho_\phi/(\sigma a)$, with maximum growth rate for $k = {\pi \hbar^2 a \over m^2 c} | g_{\phi \gamma}\dot \varrho_\phi |$, and the system tends toward to maximal helicity state \cite{Garreston-Field-Carroll-1992, Boyanovsky-1997, Field-Carroll-2000, Campanelli-Giannotti-2005}. The maximal helicity state can cause inverse cascade of the magnetic energy to larger scales \cite{Frisch-1975, Widrow-2002, Brandenburg-Subramanian-2005}.

%
%
%
\section{Discussion}
                                           \label{Discussion}

Assuming weak gravity, we formulated ED and MHD for a scalar field with general potential $V(\phi)$ and $f(\phi) F \widetilde F$-coupling. We also present ED and MHD equations for a coherently oscillating axion with $\phi^2 F \widetilde F$-coupling. The latter axion formulations use the Schr\"odinger and the hydrodynamic formulations for the axion available for the $\phi^2$-coupling. We also presented the gravitational instability of the fluid and axion caused by the MHD with helical coupling and the magnetic instability caused by the scalar field and axion.

For the QCD motivated axion with a mass around $\mu {\rm eV}$ the Jeans scale in Eq.\ (\ref{Jeans-scale}) caused by the quantum stress term is negligible. To the linear perturbation order Eq.\ (\ref{ddot-delta-axion-MHD}), ignoring the MHD contribution and the quantum stress, is the same as the pressureless matter in Eq.\ (\ref{ddot-delta-MHD}). The same is true for the nonlinear order; compare Eqs.\ (\ref{Continuity-axion}) and (\ref{Euler-axion}) with Eqs.\ (\ref{Continuity-fluid}) and (\ref{Euler-fluid-MHD-axion}). Thus, axion behaves as the cold DM. The quantum Jeans scale increases as the axion mass becomes smaller. Such axion-like particles with extremely low mass can work as a fuzzy (or wave) DM lessening the small-scale tension in the cold DM scenarios \cite{Hu-Barkana-Gruzinov-2000, Marsh-2016, Niemeyer-2020, Ferreira-2021, Hui-2021}. In this work, we call axion a massive scalar field independently of the mass and coupling to the EM field.

In the presence of the EM coupling, the Schr\"odinger and hydrodynamic formulations are {\it not} available for the conventional $\phi$-coupling. This conventional $\phi$-coupling with sufficiently large coupling strength $g_{\phi\gamma}$ can cause deviation in the DM nature of the axion, see Eq.\ (\ref{EOM-WG}). The trouble is avoided in the laboratory experiments, by {\it assuming} a sufficiently small coupling of $g_{\phi \gamma}$, which is indeed consistent with experiments \cite{Irastorza-Redondo-2018}. For example, in the experimental setting at the laboratory, with static strong aligned ${\bf B}$ and $g_{\phi \gamma}$ assumed to be sufficiently small, the generated ${\bf E}$ is small as well, thus Eqs.\ (\ref{Maxwell-1-WG}) and (\ref{Maxwell-2-WG}) give ${\bf E} = - g_{\phi \gamma} {\bf B} \cdot \nabla \phi$, and right-hand sides of Eqs.\ (\ref{Maxwell-4-WG}) and (\ref{EOM-WG}) are negligible.

For the non-negligible $g_{\phi \gamma}$ term with $\phi$-coupling in Eq.\ (\ref{EOM-WG}), however, the coherent oscillation of the axion cannot be maintained. In a perturbative sense, as the EM correction in Eq.\ (\ref{EOM-WG}) is already second-order, we can apply the Klein and Madelung transformations to the linear order, with consequent Schr\"odinger and hydrodynamic formulations. But, from the second order, the term in the right-hand side of Eq.\ (\ref{EOM-WG}) contributes, and the fuzzy (or cold) DM nature of the axion is {\it threatened}.

If the $g_{\phi \gamma}$ term with $\phi$-coupling in Eq.\ (\ref{EOM-WG}) can be ignored, we can proceed the two transformations in Eqs.\ (\ref{Klein-tr}) and (\ref{Madelung-tr}), and consequently, Eq.\ (\ref{Schrodinger-eq}) for the Schr\"odinger equation and Eqs.\ (\ref{Continuity-axion}) and (\ref{Euler-axion}) for the axion-fluid equations are valid without the ${\bf E} \cdot {\bf B}$ terms. Still, we have trouble employing the transformations in the axion-induced charge and current densities in Eq.\ (\ref{rho-j-phi}), and we have to use the field ($\phi$) instead of the wavefunction ($\psi$) or the fluid quantities ($\varrho_\phi$ and ${\bf v}_\phi$) in the ED or the MHD equations.

We can estimate the effect of axion-coupling on the MHD. In a static medium, the right-hand sides of Eqs.\ (\ref{Faraday-MHD}) and (\ref{Faraday-MHD-axion}) can be estimated as $(g_{\phi \gamma} \dot f / \sigma) c \nabla \times {\bf B}$ with $f = \hbar^2 \varrho_\phi / (2 m^2)$ for $\phi^2$-coupling. In our convention, $g_{\phi \gamma} f$, thus $g_{\phi \gamma} \dot f / \sigma$ are dimensionless. For the axion-coupling term to be important in the Faraday equation, we need the coupling constant to be $g_{\phi \gamma} \sim \sigma/\dot f$ which becomes $2 m^2 \sigma/(\hbar^2 \dot \varrho_\phi)$ for $\phi^2$-coupling. In non-relativistic fully ionized plasma, the conductivity is given as $\sigma \sim (k_B T)^{3/2}/(e^2 m_e^{1/2}) \sim 3 \times 10^{14} T_{\rm eV}^{3/2} /{\rm sec}$ with $T_{\rm eV}$ the temperature in eV unit \cite{Spitzer-1956}. Using $\dot \varrho_\phi \sim H \varrho_\phi$ with $H = 100 h {\rm km/sec/Mpc}$,we have $g_{\phi \gamma} \sim 4 \times 10^{-7} m_{22}^2 T_{\rm eV}^{3/2} /( \Omega_\phi h^3) \; {\rm cm/eV}$ where $m_{22} \equiv m c^2/(10^{-22} {\rm eV})$.

Here we note that in Eq.\ (\ref{Faraday-MHD}) a curl of misalignment between the gradient of the scalar field $\nabla \phi$ and the electric field ${\bf E}$ (for example, caused by Thomson scattering of electrons before recombination) can generate the magnetic field.

Although the coherent oscillation is preserved, the $\phi^2$-coupling is difficult to motivate in high-energy physics, and calling the case an axion may cause controversy. Despite lacking physical motivation as an axion, the successful Schr\"odinger and hydrodynamic formulations of the $\phi^2$-coupling in the MHD, structure formation, and source for $\alpha$-dynamo may deserve further study. For other couplings the Schr\"odinger and hydrodynamic formulations are not available, but we still have the ED and MHD formulations with helical coupling directly using the scalar field; see Sections \ref{SF-ED} and \ref{SF-MHD}, respectively.

%
%
%
\section*{Acknowledgments}

We thank Professors Kiwoon Choi and Dongsu Ryu, and Drs.\ Heejung Kim and Hyeonseok Seong for useful discussion. We also wish to thank an anonymous referee for constructive suggestions. H.N.\ was supported by the National Research Foundation (NRF) of Korea funded by the Korean Government (No.\ 2018R1A2B6002466 and No.\ 2021R1F1A1045515). J.H.\ was supported by IBS under the project code, IBS-R018-D1, and by the NRF of Korea funded by the Korean Government (No.\ NRF-2019R1A2C1003031).

%
%


\end{document}